# Security issues for data sharing and service interoperability in eHealth systems: the Nu.Sa. test bed

Emanuele Frontoni, Marco Baldi, Primo Zingaretti
Dipartimento di Ingegneria dell'Informazione
Università Politecnica delle Marche
Ancona, Italy
{e.frontoni; m.baldi; p.zingaretti}@univpm.it

Vincenzo Landro, Paolo Misericordia
Federazione Italiana Medici di Medicina Generale
FIMMG, Netmedica Italia
Roma, Italy
{v.landro; p.misericordia}@netmedicaitalia.it

*Abstract*— **The aim of the Nu.Sa. project is the definition of national level data standards to collect data coming from General Practitioners' Electronic Health Records and to allow secure data sharing between them. This paper introduces the Nu.Sa. framework and is mainly focused on security issues. A solution for secure data sharing and service interoperability is presented and implemented in the actual system used around Italy. The solution is strongly focused on privacy and correct data sharing with a complete set of tools devoted to authorization, encryption and decryption in a data sharing environment and a distributed architecture. The implemented system with more than one year of experiences in thousands of test cases shows a good feasibility of the approach and a future scalability in a cloud based architecture.**

*Keywords—electronic health record; security; privacy; cloud*

## I. INTRODUCTION

The need to create a set of computer science tools intended to the general medicine and governed by the same profession stems from several considerations. The first is undoubtedly the need to implement Functional Territorial Associations (FTA, AFT in the Italian acronym) of next constitution, and in general, any existing association with a tool that overcomes the difficulties to use different ambulatory software tools that do not allow a network connection among their databases, because they come from different manufacturers. In this case, in order to allow the so-called "mixed network", different databases have to be converted to a common standard so that everyone can easily interpret data. Starting from this assumption, the Netmedica Italy (NMI) company was established, with the objective to realize a digitalization and cloud computing project concerning general medicine [1], to facilitate the sharing of health data and coordinate their streams. In the work [2], the authors present a survey of new scenarios in healthcare through the emerging cloud computing technology, where many papers highlight the importance for practitioners to share healthcare data in a common system [3], [4], [5]. The authors of [6] propose a cloud computing system for sharing medical image of a hospital. The sharing of data led to the possibility of bringing to the professional sensitivity many dynamics and operations imposed by informative systems so far managed and governed by others, by undertaking a path of awareness and responsibility that the general medicine (GM), in step with the evolution of Information and Communication Technology (ICT), must certainly live as a protagonist. The term eHealth, introduced for the first time by Mitchell [7], is defined as the use of ICT across the whole range of healthcare functions and can contribute to cope with the challenges currently faced by healthcare systems in Europe. These challenges include the need to ensure system sustainability while preserving quality in the face of an ageing population. The adoption of eHealth, matched by organisational changes and by other technical innovations, can turn these challenges into the triple wins set as targets of the European Innovation Partnership on Active and Healthy Ageing (EIP AHA): quality of life, sustainability, innovation and growth. There are many recent works that put in evidence the importance of confidentiality for eHealth data systems, concentrating on the security framework [8], [9], [10] and in [11] the reasons for maintaining medical data private and restricting the access are explained. According to [12], security and privacy for data sharing can be ensured if technology, legal and social aspects are considered during the realization of eHealth systems. They retain that departing from a sociotechnical analysis, security and privacy in eHealth are possible.

In the general Italian medicine scenario, the adoption of eHealth in primary care by General Practitioners (GPs) is pivotal to realise the above mentioned potential.

We believe that the development of a common platform devoted to GPs allows to "create a system" which, besides to give an image of compactness (and also promote the instances of representativeness and "contractual strength"), provides a basis for the endless opportunities of interoperability with all other available databases. In this context, each practitioner might already be ready to share subsets of health and care data of his patients with other practitioners. It is possible to imagine a system in which the same GP accredits the patient to access a protected part of his clinical record, where activities of informative collaboration, authorization or denial for sharing of health information, acquisition by the patient of clinical data to make them available to other doctors for consultation are allowed. It is also appropriate and technically possible, having to manage a single database in the cloud, to encourage the

This work was supported in part by the MIUR project "ESCAPADE" (Grant RBFR105NLC) under the "FIRB – Futuro in Ricerca 2010" funding program.

doctor in the process of auditing and reporting, equipping him with the evolved tools for database querying; as it is possible to support the doctor with professional services aimed at improving and facilitating his activities. Federsanità Anci, with its eHealth sector, fully supports goals and intentions proposed by NMI and, supposing to create a deeper integration with other databases of healthcare systems, it has strongly shared the initiative, by participating in the establishment of the Foundation NU.SA. - Cloud Health. Nu.Sa. (Nuvola Sanitaria) is an Italian project going in that direction and born in Italy in 2012.

## II. THE NU.SA. PROJECT

The aim of the Nu.Sa. project is the definition of national level data standards to collect data coming from GP's Electronic Health Records (EHRs) on a cloud based architecture. Today, the project involves thousands of Italian GPs collecting data from millions of patients. Two key features of the Nu.Sa. system are: the referral of patients and portability of data. In most clinical environments around the world and also in Italy, the healthcare of the individual patient is shared between Specialists and GPs. Often there is a gap in the exchange of important information between the two groups and Nu.Sa. can provide a solution at a national and European level to cope with this limitation.

Data are collected and shared using an open standard [13] and the system architecture is mainly focused on data sharing and interoperability with an intensive effort on privacy and data security. This paper is mainly based on the description of the security concerns relevant to these systems and services, and on the technical solutions, which are most suitable to overcome them.

In fact, while there are some international standards, which regulate the main requisites of eHealth data systems in relation to consumers, like the ASTM E2211 standard[14], security and privacy issues are regulated at a national level, and therefore must be addressed by taking into account the national regulations.
By focusing on the case of Italian GPs, the most important security requisites are: i) to guarantee that data are properly collected and handled, ii) to guarantee that data are safely stored and storage is limited within the Italian territory, iii) to guarantee that data are accessed only by authorized doctors and that iii) data integrity and consistency is ensured..

### A. Electronic health record

The EHR (denoted by the Italian acronym FSE, or *Fascicolo Sanitario Elettronico*) is a collection of electronic documents available to doctors, pharmacists and hospitals, which collect the health data of a patient, such as diseases, surgical procedures, medical tests, prescribed medications and hospitalizations. In 2008, with a view to modernizing the health care system, both public and private, it has been established with the aim of creating a relationship between health care professionals and organizations involved in the life of the individual through clinical documents which are continuously updated using a computer system. The purposed of the FSE are prevention, diagnosis, treatment and rehabilitation of the patient who can freely choose whether or not to converge clinical information concerning him in this electronic file, ensuring also the possibility that his health information will remain available only to authorized professionals or sanitary organisms. In case the patient does not give the consensus, since it is completely optional, it cannot proceed with the treatment, but it has not any effect on other consensus to other treatments provided in health care. The D.L. n. 69 of 2013, so-called Decree of the Making (or *Decreto del fare*, in Italian), specified the role of the Agency for digital Italy, which hosts project plans presented by the Regions and Autonomous Provinces for the implementation of the FSE and establishing the criteria for the creation no later than 30 June 2014.

In the context of cooperation among health care facilities is expected that every structure creates its Electronic Health Dossier (*Dossier Sanitario Elettronico*, DSE, in Italian), all the DSE will form the FSE.

### B. Functional territorial associations

In recent years, the primary health care of the citizens has faced radical changes in order to improve the efficiency and capability of providing public medical assistance to patients. The law n. 189/2012 (Balduzzi) with the national collective agreement (ACN) provides a reorganization of primary care in mono-professional organizational forms called functional territorial associations (FTA, Art. 1), complemented by the Complex Primary Care Units (UCCP). On this side, ICT is considered a fundamental tool for functional aggregation and integration of local and hospital care. In this context, the production of a report to support the planning and control of FTA activities plays an important role.

However, in the text of the law 189/2012, the modality of healthcare data treatment in a computerized way is not cited, neither their safety nor their access, demanding these issues to the guidelines introduced in the D. L. 196/2003 regarding personal data security.

## III. SECURITY AND PRIVACY ASSURING TECHNIQUES

The use of robust and up-to-date information security and privacy techniques is a fundamental pillar of the Nu.Sa. project. There are two main requirements to fulfill:
1. Avoid unauthorized access to the healthcare data of any patient (*security requirement* or *confidentiality requirement*).
2. Avoid that authorized users other than the medical doctor who is in charge of a patient are able to match the name of that patient with his healthcare data (*privacy requirement*).

These two requirements must be fulfilled by exploiting suitable design solutions for the hardware/software architecture and the encryption of healthcare data. They represent two challenges which must be faced together, since the design choices affect the choice of the encryption techniques and vice-versa.

Concerning the privacy requirement, an intriguing solution is represented by the use of the so-called *negative databases* [15], in which the information is complemented before being stored. In other terms, all the strings which do not describe the identity of a user are stored in the database, such that a search query on that user is successful only if it returns no matches. Despite this is an ingenious way of protecting the users' privacy, it has several drawbacks, since it considerably increases the storage space and the management complexity.

Therefore, we consider a different approach which simplifies the practical implementation of the system.

Another solution to provide for users' data privacy in databases is represented by the concept of *k*-anonymity and its variants [16], [17], [18]. The rationale of this approach is to analyze the sets of attributes of each user and to store in each database only a subset of them which are common to a sufficiently large set of users, in such a way as to introduce some ambiguity if one tries to identify the user from such a set of attributes. This approach has the advantage of not requiring the use of cryptographic functions to protect privacy. However, it requires to perform a thorough analysis of the common attributes among users both the first time the database is populated and each time it is updated.

In the considered scenario, we need to use cryptographic functions to provide for data confidentiality, and we can also exploit some peculiar features of the existing hardware/software architecture. This allows us to achieve the desired levels of security and privacy without the need to resort to negative or *k*-anonymous databases.

The block scheme of the architecture we use to achieve the desired level of security and privacy of the users' data is reported in Fig. 1. The main components of the system are:

- In the EHRs, each patient is identified through a unique, randomly generated Patient Identifier (PID). Then, for each patient the EHRs contain the records and files collecting his diseases, clinical exam reports, treatments, previous surgeries, etc.
- The Patient Registry (PR), which contains a personal data record for each patient registered to the service. This personal data record includes the identifiers of the Medical Doctors (MDs) who are in charge of a patient and, for each MD, an Enciphered Patient Identifier (EPID).
- The Aggregation and Login Server (ALS), which provides a frontend for the MD terminals and for the extraction, manipulation and presentation of the users' data.
- The MD terminals: each MD is expected to have a master terminal (*e.g.*, a workstation at his office) and a set of slave terminals (*e.g.*, his mobile phones, tablets, notebooks, etc.).
- The patient terminal: the system can allow patients to access the system.

This architecture is able to achieve the security and privacy targets by exploiting a basic fact which characterizes the current scenario: each MD has a master terminal in which the healthcare data of the patients he is in care of are stored and associated to the patients themselves. This situation stems from legacy software tools which are installed at the MDs' offices, and contain a local copy of the medical records concerning their patients. The MDs are allowed to associate the identity of each patient with his medical record, and the security of these data is protected through classical methods (like operating system passwords and local database encryption).

Based on such a scenario, the following procedures are used to populate, update and query the online system depicted in Fig. 1.

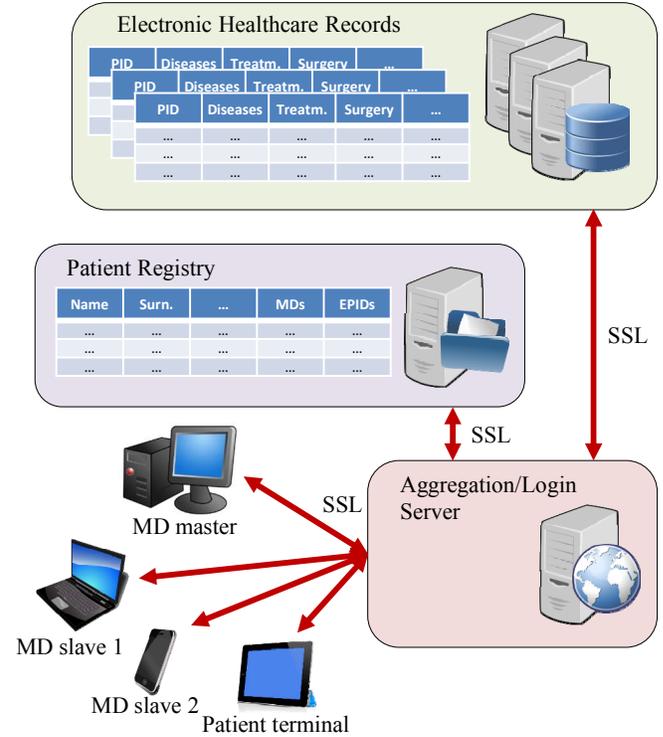

Fig. 1. Security and privacy assuring architecture.

*A. Database population*

When a MD subscribes to the service, he has the ability to populate the online registries with the personal and healthcare data of the patients he is in care of. For this purpose:

1. The MD authenticates into the system through the ALS by using his master terminal.
2. The local software client generates a unique PID for each patient having a record in the local database, and stores the PIDs in the local database too.
3. The local software client uses a symmetric encryption algorithm ($E$) and a private key owned by the MD ($K$) to obtain an Encrypted PID ($EPID_K$) for each PID, *i.e.*, $EPID_K = E_K(PID)$. Details on the encryption algorithm will be provided in the next section.
4. The ALS receives a record containing the personal data of the patient, the name of the MD and the $EPID_K$, and creates a new entry in the PR to store these data.
5. The ALS receives a record containing the PID of the patient and his medical data, and creates a new entry in one or more EHRs to store these data.
6. Some EHRs may be already populated with the patient data (e.g., when they are hosted by the healthcare center where the data was created). In this case, the web services exposed by the HR are used to query it based on its own indexing, and the PID is added to the matching records and data.

*B. Patient query*

When a MD wishes to perform a query concerning one of his patients, he works from any of his terminals (master or slaves), and:

1. He authenticates from the local software client into the online system through the ALS.
2. He performs a query to the ALS based on the name (or other personal data) of the patient.
3. The ALS queries the PR, checks that the patient is associated to the MD and return the corresponding $EPID_K$.
4. The local software client uses a symmetric decryption algorithm ($D$) and the private key owned by the MD ($K$) to obtain the PID from the $EPID_K$, *i.e.*, PID = $D_K(EPID_K)$, and queries the ALS with such a PID.
5. The ALS queries the EHRs with the PID and presents the medical data of the patient to the MD terminal.

*C. MD roistering*

An important feature of this system is to support roistering of the MDs. This is necessary, for example, in functional territorial associations of MDs, where more than one MD assists a patient, or when a MD is out of work and must be temporarily substituted by another MD. In order to cope with these cases, the system exploits a feature, which already exists in the Italian healthcare system: each patient is associated to a Primary MD (PMD), who is administratively responsible for the healthcare of that patient. Then, one or more Secondary MDs (SMDs) may temporarily or permanently substitute the PMD. The system exploits this fact and relies on the PMD for enabling the patient data access by SMDs. In order to enable a SMD:

1. The PMD authenticates into the system through the ALS from one of his terminals (master or slaves).
2. The PMD performs a query to the PR, through the ALS, based on the personal data of the patients for whom he wishes to enable the SMD.
3. The ALS retrieves the list of the EPIDs obtained from the PR for those patients, and sends them (or puts them in a sending queue) to the SMD local client.
4. The first time the designated SMD authenticates into the system, he is asked to accept the designation sent by the PMD. After accepting, the SMD receives the list of EPIDs and encrypts again each of them by using his private key ($SK$), *i.e.*, $EEPID_{K,SK} = E_{SK}(EPID_K)$. Then he sends them back to the ALS.
5. The ALS receives the doubly encrypted PIDs and sends them (or puts them in a sending queue) to the PMD.
6. The next time the PMD authenticates into the system, he receives the notification that the SMD has accepted his designation, together with the list of the doubly encrypted PIDs. He then decrypts each of them with his private key ($K$): $EPID_{SK} = D_K(EEPID_{K,SK})$, thus obtaining a list of PIDs encrypted through the SMD private key. The cryptographic functions to be used for this purpose are described in the next section.
7. For each patient in the list, the PMD sends the corresponding $EPID_{SK}$ to the ALS, which adds it to the corresponding record in the PR, together with the name of the designated SMD.

This way, after authenticating into the system through the ALS, the SMD can perform a query concerning the patients for whom he has been delegated, by following the same procedure described in Section IV.B. The SMD does not need any master client, since he is not administratively responsible for the medical records of such patients.

Concerning the revocation of the PMD designations, it can be easily implemented by including in the PR a field, which contains the time intervals in which each a pair (SMD, EPID) is valid. This way, each SMD can be delegated for a limited period, or only at some specific times during each day, week or month. Then, some background task has to be periodically run to clear the expired entries from the PR. Obviously, the PMD can also manually ask the system to remove the pairs (SMD, EPID) which he wants to revoke.

*D. Medical records updating/replacement/removal*

A necessary task is to periodically update, replace or remove some patients' medical records. This can be easily done through the architecture described in Fig. 1. In fact, both the PMD and the SMDs are allowed to write in some of all EHRs, through the ALS. This way, they can update or replace the patients' medical records, if they wish. All the terminals will then see the updated medical records each time they query the updated EHRs. A background task must be periodically performed to synchronize the local database in the PMD master terminal with the contents of the updatable EHRs.

A different situation concerns the insertion of new patients and the removal of registered patients. These tasks are only allowed by the PMD who is responsible for those patients. If he wishes to add a new patient to the system, he simply has to follow the procedure described in Section III.A. Then, he will be automatically associated to the new patient as his PMD. If instead the PMD wishes to remove a patient, he can execute such a task through the ALS. This is performed in two stages: *i)* the PMD sends the PID to the ALS, and the ALS removes the corresponding record from the editable EHRs, *ii)* the PMD sends the EPID to the ALS, and the ALS removes the corresponding entry from the PR, including the EPIDs of the possible SMDs of that patient. The operations of insertion and removal of patients should be performed from the PMD master terminal, in such a way as to keep the local database updated. Alternatively, these procedures can also be enabled from the PMD slave terminals. However, in such a case, the slave terminals must store local databases as well, and a parallel system must be deployed to periodically flush them to the local database in the PMD master terminal.

A particular situation arises if the PMD loses his private key. In such a case, he is no longer able to decrypt the EPIDs he computed, and therefore the system is unable to retrieve from the EHRs the medical records of his patients. This is a drawback of this solution with respect to more involved approaches, like negative databases and *k*-anonymity. However, we can overcome this problem by exploiting the fact that the PMD has a master terminal, which contains the medical records of his patients and their PIDs. Therefore, if the PMD loses his private key, he can generate another one, and regenerate the corresponding EPIDs from his master terminal. Then, the new EPIDs must be sent to the ALS, which replaces them into the PR. If instead a SMD loses his secret key, he must notify the system, which then requests the PMD to delegate the SMD again, by following the procedure described in Section IV.C. This allows the SMD to re-compute his associated EPIDs with his newly generated private key, and the ALS then replaces them into the PR.

*E. Data obfuscation*

In some cases, although the patient identity is unassociated to his medical data contained in the EHRs, it may be requested to obfuscate the latter. For example, this may occur for some text fields where the MD can insert free notes, which may inadvertently identify the patient. For this reason, the system must provide the chance to obfuscate some of the data contained in the EHRs.

This is achieved by using the same symmetric encryption algorithm ($E$) used for computing the EPID starting from the PID, whose details are provided in the next section. However, in this case, the secret key used for encrypting the data to obfuscate is not the MD secret key ($K$), but is computed starting from a combination of the patient's personal data contained in the PR. This way, any MD (the PMD or the delegated SMDs) who is able to match the PID with the patient identity contained in the PR is also able to compute the secret key needed to decrypt the obfuscated data. Concatenating some of the personal data of the patient, extracted from the PR, and then computing a hash digest of the resulting string obtain the secret key.

Using the secret key so computed has the advantage of avoiding the replication of the obfuscated data which would be necessary by using the PMD and SMDs secret keys. On the other hand, a drawback of this approach is that, since the algorithm used for computing the secret key is known, a malicious user having access to the PR and EHRs could compute all the secret keys corresponding to all the patients and try to decrypt some data contained in the EHRs. This way, if he finds a match, he is able to discover the corresponding patient identity and to decrypt his obfuscated data. However, in order to carry out such an attack, the attacker must have access to both the PR and the EHRs. In addition, the computation of a patient's secret key can be made harder by exploiting a proof-of-work scheme [19], such that testing very large numbers of keys becomes practically infeasible.

*F. Patient access*

The system provide patients with the chance to access their medical records, and optionally to make some choices concerning them. In order to allow a patient to access his medical data, the system exploits a procedure which is very similar to that used for delegating a SMD.

When a patient wished to access his data, he must subscribe to the service through the ALS. Then, the ALS queries the PR with the patient identity and finds his corresponding EPID encrypted with the PMD secret key. The ALS sends the EPID to the patient, who encrypts it again with his secret key before sending it back to the ALS. The next time the PMD logs into the system, he finds the pending request from his patient. If he accepts them, the ALS sends the EEPID to the PMD local client, which decrypts it with the PMD secret key. This way, an EPID encrypted with the patient secret key is obtained, and it is stored in the PR as well, such that the patient can retrieve his medical data in the same way as for MDs. An alternative approach would be to provide each patient with is PID in clear, but it is preferable to avoid the unessential circulation of the PID, especially into devices where it could be associated with the patient identity (like the patient's terminals).

After accessing the system, a patient may have the ability to decide which obfuscated medical data are visible to which MDs. If the access to an obfuscated data record is disabled for one or more MDs assisting a patient, the ALS will prevent those MDs from retrieving the obfuscated from the EHRs.

IV. DATA ENCRYPTION, MANIPULATION AND RETRIEVAL

As it results from the descriptions provided in the next sections, the proposed system requires an intensive use of data encryption and manipulation algorithms.

First of all, the PIDs and some of the medical data must be encrypted. For this purpose, our system uses up-to-date symmetric encryption schemes, like AES [20]. A special feature, which is needed by the symmetric encryption scheme adopted by the system, is the chance to revert the decryption order. In fact, in order to manage the MD rostering (see Section III.C), it is necessary that $EPID_{SK} = D_K(EEPID_{K,SK})$, where $EEPID_{K,SK} = E_{SK}(EPID_K)$. Therefore, two successive encryptions and decryptions must work independently of the order in which the corresponding secret keys are used. This can be implemented by using a symmetric cipher (like AES) in Output FeedBack (OFB) or Counter (CTR) mode [21]. A drawback of these two operating modes is that, if an attacker intercepts the plaintext and its corresponding ciphertext (*i.e.*, the EPID and the EEPID), he is able to forge encrypted messages. However, in the considered system, an attacker would not be able to generate a valid EPID to encrypt a second time, therefore he is unable to break the system privacy.

Concerning the security of the databases containing the PR and EHRs, in a classical centralized architecture (a few databases stored in a few servers) a possible solution consists of using the Transparent Data Encryption (TDE) function provided by several database engines. This way, even if a copy of the whole database is stolen, the data it contains is protected from unauthorized accesses. Moving towards a cloud architecture, a promising solution consists of exploiting the decentralized nature of the storage space and the intrinsic data fragmentation to achieve security. In fact, security can be obtained by using coding together with all-or-nothing transforms and data dispersal [22]. This way, each storage node only possesses a small fraction of the whole archive, and cannot retrieve any information from it. Furthermore, this allows to protect the sensitive data from possible malicious cloud service providers or compromised storage nodes.

Another issue concerns the chance to perform search queries on the personal and medical data of the patients. In the architecture illustrated in Fig. 1 and described in the previous section, if a user has the rights to access the PR or the EHRs, he can perform searches on the data stored in clear (*i.e.*, without obfuscation). Concerning the obfuscated data, the simplest solution is that the authorized user client retrieves them, decrypts them and then performs the search in the decrypted domain. This solution, however, requires large bandwidth and storage space, and may be intolerably slow when working on resource-limited devices. Another solution can be to build an index for each obfuscated data entry, containing a small set of keywords which are then stored in clear. This way, the search queries on the obfuscated data are accelerated, at the cost of some storage overhead and processing power at the time of data entry. A third, pioneer solution consists of implementing a searchable encryption scheme [23], which allows to perform searches directly in the encrypted domain.

Finally, another important function to be provided by the system concerns the computation and extraction of statistical data from the PR and EHRs. This is a very important outcome of the implementation of a medical information system of the type considered. This task can be easily accomplished on the non- obfuscated data, possibly by exploiting the processing power provided by the cloud infrastructure, thus avoiding to transfer the computational burden to the client terminal. Concerning the obfuscated data, a first approach consists again of retrieving them locally, decrypting them and performing the due computations. This however once again has the drawback of requiring large bandwidth, storage space and local computing power. Another solution would be to exploit homomorphic encryption schemes [24], [25], which allow to perform computations in the encrypted domain, and thus to outsource computations (*e.g.*, to cloud service providers). Although a lot of research work on homomorphic encryption schemes is still ongoing, some practical solutions already exist. For example, the system proposed in [26] can compute some statistical parameters, like means and variances, in the encrypted domain.

## V. Conclusion

We have described the Nu.Sa. project, together with its state of the art and future goals. The aim of the Nu.Sa. project is the definition of national level data standards to collect data coming from General Practitioners' Electronic Health Records and to allow secure data sharing between them. We have presented a software architecture which is able to achieve the desired levels of security and privacy of the patients' data, by leveraging suitable cryptographic techniques, while limiting complexity and allowing a great level of scalability and interoperability.